\documentstyle[prl,aps,twocolumn,psfig,floats]{revtex}
\input{epsf}
\begin{document}
\draft
\twocolumn[\hsize\textwidth\columnwidth\hsize\csname@twocolumnfalse\endcsname

\title{Vortices Freeze like Window Glass: the Vortex Molasses Scenario}
\author{Charles Reichhardt$^1$, Anne van Otterlo$^2$ 
and Gergely T. Zim\'{a}nyi$^1$}
\address{$^1$ Physics Department, University of California, 
Davis, CA 95616, USA}
\address{$^2$ Instituut-Lorentz for Theoretical
Physics, P.O.B. 9506, 2300 RA Leiden, The Netherlands}
\date{\today}
\maketitle

\begin{abstract}
We overview several recent experimental and numerical observations, which
are at odds with the Vortex Glass theory of the freezing of disordered
vortex matter. To reinvestigate the issue, we performed numerical simulations
of the overdamped London - Langevin model, and use finite size scaling
to analyze the data. Upon approaching
the transition the initial Vortex Glass type criticality is arrested
at some crossover temperature. Below this temperature the timescales 
continue growing very quickly, consistent with the Vogel-Fulcher form, 
while the spatial correlation length $\xi$ stops exhibiting any 
observable divergence. We call this mode of freezing the Vortex Molasses
scenario.
\end{abstract}
]

\begin{narrowtext}

The influence of disorder on vortex matter is one of the most paradigmatic problems.
The vortex lattice, formed in clean systems, is inherently unstable towards 
a less ordered state even for infinitesimally small disorder \cite{larkin}.
At small magnetic fields, or equivalently, at weak disorder, a dislocation free 
phase is emerges, which thus retains a topological order.  Small angle neutron 
scattering \cite{forgan93}, and Bitter decoration experiments \cite{bishop} 
seem to support this picture.  The theoretical foundation for such a Bragg Glass
was provided by scaling arguments~\cite{nattermann}, and variational 
calculations \cite{giam1}.  Its cornerstone is the logarithmic behaviour
of vortex correlations at large distances.  Numerical simulations also reported a 
strongly suppressed dislocation density \cite{ryu96}, and confirmed the logarithmic 
behaviour of correlations \cite{us1} below a critical field strength. 
It is noteworthy, however, that the largest scale imaging studies \cite{lieber},
did not find evidence for logarithmic correlations, thus the 
details of the dislocation free regime are still subject to discussion.

We also understand the influence of increasing fields, or disorder. The key 
phenomenon here is the appearance of dislocation loops, accompanied by the 
entanglement of vortices \cite{ertas,giam2}.  Experimental support for this idea
is the sharp enhancement of the critical current from magnetization measurements
\cite{hardy94,klein94,zeldov96,groot97}, the rapid destruction of the Bragg 
peaks in neutron scattering \cite{forgan93}, and the pronounced dips in the 
electric field - current density, or E-J
curves~\cite{safar95}. Numerical studies found evidence for dislocation loops 
destroying a quasi-ordered state in frustrated XY models~\cite{gingras}, 
in Lawrence-Doniach representations~\cite{ryu96}, and in realistic
London-Langevin approaches \cite{us1}.

The nature of the high field phase is still very much in debate.
The thermally assisted flux flow (TAFF) picture predicts
that vortices move in bundles, and overcome barriers via thermal
excitations. This destroys superconductivity
because the linear resistivity assumes a finite value, governed by an
activated temperature dependence $R(T) \sim R_{0}~{\rm exp}(-U/T)$ \cite{kim}.  
An influential alternative was put forward in the form of the Vortex Glass (VG) 
theory \cite{fisher89,ffh}.  The proposed Vortex Glass phase is 
distinguished by an unbounded distribution of barrier heights.
This results in the vanishing of the linear resistivity, thus restoring
superconductivity, and inherently non-linear E-J characteristics. Numerical support
for this picture emerged from the study of the isotropic Gauge Glass model, ignoring
the effects of screening \cite{young1}. Experimental confirmation 
soon followed, on heavily twinned YBCO films \cite{koch}. A key evidence
was provided by observing the scaling of a crossover current $J_{x}$ \cite{gammel}. 

Recent experimental and numerical work, however, has raised new questions 
about the Vortex Glass picture. 

$i)$ The values of the correlation length, creep and dynamical exponents, 
$\nu, \mu$ and $z$, respectively, 
seem to depend on temperature, current, 
and sample quality \cite{melissa} in a very nonuniversal way.  
$\nu$ was found between $1.3-2$, $z$ between
$3.1-6.5$, and $\mu$ between $0.2 - 0.5$. 

$ii)$ The above values of $\nu$ and $z$ are much higher than their
mean field values, indicating that the lower critical dimension might be close 
to 3. It is already accepted that there is no finite temperature 
VG phase in 2D \cite{young1}.

$iii)$ Recent experiments in completely untwinned YBCO samples found
that the $E-J$ curves remained completely linear down to the lowest measurable
values of the current. Correspondingly no scaling behaviour of the $E-J$
curves were found \cite{crabtree,yeh}. This suggests that in previous works
the twin boundaries might have played the role of extended defects, 
and in fact the observed
scaling behaviour was that of the Bose Glass. 

$iv)$ When the twin boundaries were removed in YBCO samples, the crucial 
crossover current $J_{x}$ was found to saturate, instead of exhibiting
a scaling behaviour \cite{lopez}.

$v)$ Recent numerical works reported that when a finite London screening
length $\lambda$ was restored into the previously studied Gauge Glass models,
the finite temperature Vortex Glass transition disappeared 
\cite{young97,kawamura,rieger}. Now, close to the transition
the vortex correlation length is supposed to diverge, thus
exceeding $\lambda$. Therefore the ultimate transition region is always 
in this finite $\lambda$ regime.

We conclude that there is an emerging body of evidence, which is
inconsistent with the Vortex Glass picture. Motivated by this inconsistency, 
in this paper we explore analogies to an other widely studied glass transition, 
and investigate the possibility that {\it Vortices freeze like the window glass: 
the Vortex Molasses scenario.}

We realize that there is not a uniquely accepted theory of the
window glass transition. Therefore we construct the Vortex Molasses (VM) scenario
only from those elements, which {\it are} common among the different theories: 
$i)$ a very rapid freezing of the dynamics, with diverging timescales, 
characterized by the Vogel - Fulcher law: $\tau \sim {\rm exp}[1/(T-T_G)]$; 
$ii)$ the possible divergence of the spatial correlation length $\xi$ is 
rendered unobservable by this rapid freezing.

We note that among the early alternative propositions, 
some emphasize the entanglement of vortices in the presence of disorder, 
in analogy to polymer glasses \cite{marchetti}. Also, extensions of the TAFF 
theory were constructed \cite{littlewood}. Finally, a Vortex Slush picture 
has been proposed, viewing the glass of vortices as a viscous liquid,
driven by the remnants of the first order melting transition \cite{slush}. 

We start by overviewing ways to distinguish between the VG and VM scenarios.
First, the predicted temperature dependence of the resistivity
differs: in the VG theory the resistivity
vanishes as $\rho(T) = \rho_{VG} |T-T_{G}|^{\nu(z-1)}$, whereas in the VM 
scenario one expects the resistivity to follow the Vogel-Fulcher law 
$\rho(T) = \rho_{VM} {\rm exp} [-1/(T-T_G)]$ \cite{goetzebook}.
However it is hard to achieve decisive distinction between these forms,
as the resistivity exponent in the VG theory is large: $\nu(z-1) \sim 5 - 7$
\cite{gammel}.

The second method is more promising. Equating the current related free energy 
of a correlated volume to the thermal energy yields the above mentioned 
crossover current density scale $J_x = 
c k_{B}T/(\Phi_{0}\xi^{2})$ \cite{ffh}. Above the transition temperature
the low current linear $E-J$ of the viscous
liquid is expected to cross over at $J_x$ to the Bardeen - Stephen form at high 
currents. In the VG theory the correlation length $\xi$ diverges: 
$\xi \sim (T-T_G)^{-\nu}$. Correspondingly the crossover current scale $J_x$ 
collapses as $J_x(T) \sim (T-T_G)^{2\nu}$. In contrast, in the VM picture 
the criticality of the correlation length is {\it unobservable}, 
thus $J_x$ {\it does not
collapse}. As mentioned, while in twinned samples $J_x$ collapsed, in
untwinned YBCO $J_x$ {\it saturated} at some finite value upon approaching
the transition \cite{lopez}.

In this paper we report numerical simulations of a realistic, London-type model
for driven vortices, governed by overdamped dynamics in order to distinguish
between the above two scenarios. Previous studies on the gauge glass
already indicated a breakdown of the VG picture 
\cite{young97,kawamura,rieger}, however that model 
is rather simplified. For instance it is isotropic, whereas in real vortex 
matter the external field definitely introduces a strong anisotropy.
Thus it remains an open question, whether the gauge glass
adequately describes the vortex matter, making our realistic simulations necessary.
The Langevin equation describing the overdamped motion of vortices reads
\begin{equation}
  \eta\frac{\partial{\bf R}_{\mu}(z,t)}{\partial t}=
  {\bf \zeta}_{\mu}(z,t)+{\bf F}_{L} -\frac{\delta H[\{{\bf
      R}_{\nu}(z,t)\}]} {\delta {\bf R}_{\mu}(z,t)}\;,
\end{equation}
where $\mu$ labels the vortices with coordinates ${\bf R_{\mu}}$,
$\zeta(z,t)$ is the Langevin noise, and ${\bf F}_{L}$ is the Lorentz force.
The Hamiltonian $H$ is constructed on the basis of the
London theory. Its derivative decomposes into three forces:
the pairwise interactions, the single-vortex bending force, and the pinning.
For more details of the method, see Ref.\cite{us1}.
The trustworthiness of our code was demonstrated by the quantitatively
correct reproduction of the phase diagram of disordered YBCO \cite{us1}.
In our model we do not take into account vortex-loops whose effect on 
melting is still controversial \cite{Sudbo}. 

There are several length scales in the model. To avoid observing some 
crossover instead of the asymptotic behaviour, 
we chose the characteristic microscopic length scales small 
and close to each other. We use $\lambda/\xi = 4$, a large magnetic field of 
$H/H_{c2}=0.2$, to make $a_{0}\approx \lambda$, and finally we made the system 
isotropic by choosing $\epsilon = 1$. The 100-500 vortex elements produce good 
self-averaging, so a reasonable statistics was achieved by averaging 
over 10-20 disorder realizations. Simulated annealing
was employed to generate the starting configurations.

In Fig.1 we show the typical behaviour of the differential resistivity 
$\rho(T,I)=dE/dJ$, normalized to $\rho_{BS}$, the Bardeen-Stephen value. At high 
currents $\rho(T,J) = \rho_{BS}$ as it should. 
After a pronounced drop with 
decreasing current, $\rho(T,J)$ flattens at low J, 
clearly indicating an ohmic 
behaviour: we are in the Vortex Liquid regime.

\begin{figure}[t] \unitlength1cm
\centerline{
\epsfxsize=8.0cm
\epsfysize=6.5cm
\epsfbox{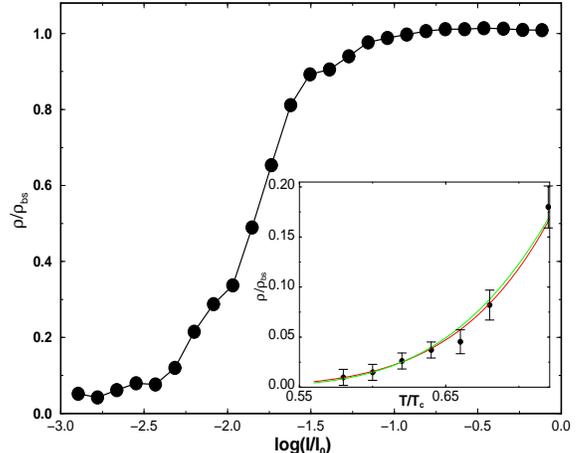}}
\caption{The current dependence of the resistivity $\rho(T,I)/\rho_{BS}$ 
at $T/T_{c} = 0.74$. Inset:
The temperature dependence of the low - current ohmic resistance 
$\rho(T,I->0)/\rho_{BS}$, for $L=8^3$ systems. Dark gray: fit to the Vogel - Fulcher
form: $\rho(T)=\rho_0 {\rm exp}[-T_0/(T-T_G)]$, 
light grey: VG fit: $\rho(T)=\rho_0 [(T-T_G)/T_G]^{2\nu}$. 
}
\end{figure}

In our temperature sweeps $\rho(T)$ drops by two orders of magnitude 
upon approaching the freezing transition.
The inset of Fig.1 shows its temperature dependence. We also exhibit
a Vogel - Fulcher fit, and a power law fit. As expected, both fits are
comparable, and thus do not distinguish between the VM and VG theories.


However, useful information can be extracted from the $R(T)$ runs via
finite size scaling. In Fig.3 we show the results for $4^3, 6^3,$ and $8^3$ systems.
Upon cooling, the system exhibits increasing finite size sensitivity down
to $T/T_c \sim 0.70$, which typically indicates the approaching of a
phase transition. Remarkably, however, below this temperature range
the finite size sensitivity decreases, as if the criticality is arrested.

\begin{figure}[t] \unitlength1cm
\centerline{
\epsfxsize=8.0cm
\epsfysize=6.5cm
\epsfbox{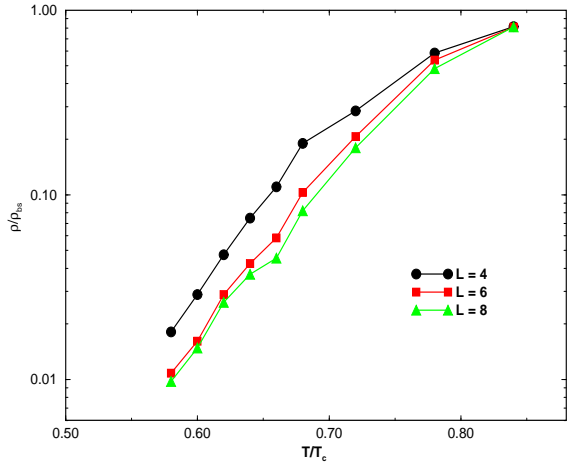}}
\caption{The temperature dependence of $R(T)$ for systems with $L=4,6,8$.}
\end{figure}

To view this from a different perspective, we follow Young et al. \cite{young97}
by studying an analogue of the Binder - ratio, $log [\rho(L)/\rho(L')]/log[L/L']$.
Far from a transition, where finite size sensitivity is small, this should be
close to $0$. Approaching a transition the increased finite size sensitivity is
signalled by the data on different system sizes splaying out. Eventually, however, 
they come together and {\it cross at} $T=T_G$. In Fig.4 we show this ratio for our 
model. As the temperature decreases, the initial splaying shows an 
impending transition. The curves, however, do not cross. Instead, they turn back 
up: the transition is arrested. This again signals the decrease of finite
size sensitivity, which is most readily interpreted as the initial increase
of the correlation length $\xi$ being arrested around $T/T_c \sim 0.70$.

\begin{figure}[t] \unitlength1cm
\centerline{
\epsfxsize=8.0cm
\epsfysize=7.0cm
\epsfbox{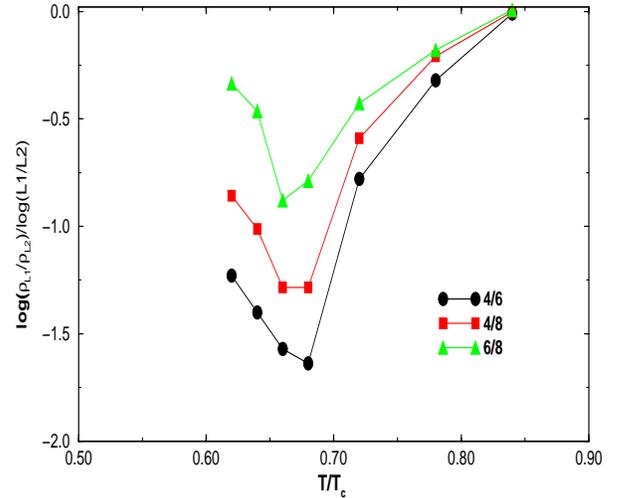}}
\caption{The logarithmic resistance ratio as a function of the temperature.}
\vspace{-1cm}
\end{figure}

The best measure of $\xi$ is via the crossover current $J_x$, plotted
in Fig.5. As $\xi \sim J_x^{-1/2}$, decreasing $J_x$ indicates increasing 
correlation length. However $J_x$, and thus $\xi$, 
{\it saturates} with decreasing $T$, around
$T/T_c \sim 0.68$, in quantitative accord with the finite size scaling.
Note that a very similar flattening of $J_x$ was
observed in untwinned YBCO \cite{lopez}.
All these three tests can be interpreted as follows. Upon decreasing
the temperature a Vortex Glass criticality starts to develop. However
this critical behaviour gets arrested around $T/T_c =0.69 \pm 0.01$, and
crosses over to a Vortex Molasses criticality. This is characterized
by a rapid, Vogel - Fulcher type decrease of the resistivity, but at the same 
time an essentially noncritical behaviour of the correlation length $\xi$.

\begin{figure}[t] \unitlength1cm
\centerline{
\epsfxsize=8.0cm
\epsfysize=6.5cm
\epsfbox{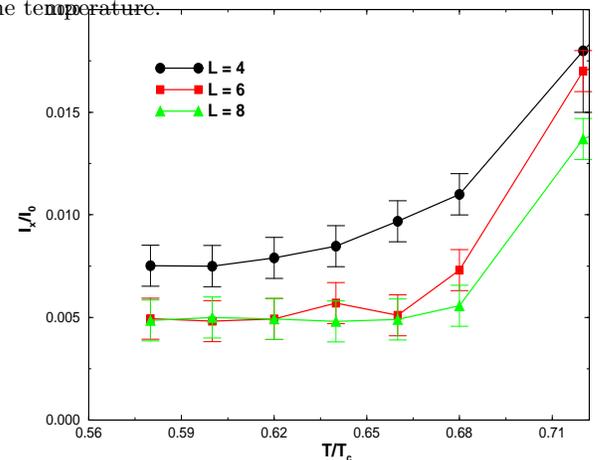}}
\caption{The temperature dependence of the crossover current $I_{x}$, for $L=4,6,8$.}
\end{figure}

The above results established that the freezing transition is unlikely 
to be governed by Vortex Glass theory, but rather it looks more like a
window glass transition. However there isn't a single theory agreed upon
by the window glass community. For a review of different approaches, 
see Ref. \cite{goetzebook}. Some theories propose that the correlation
length diverges as a power law, but the freezing of the dynamics is so
rapid, that it renders this divergence unobservable. Others believe that in fact
$\xi$ does not diverge at all, it remains noncritical even on the longest 
time scales. Finally there are theories which envision that there is no
true transition at any finite temperatures, but a rapid, continuous increase
of the viscosity, diverging only at $T=0$. This latter view was imported
to the vortex problem by Ref. \cite{young97}. Setting up the scaling theory
for finite size systems with a correlation length $\xi \sim T^{-\nu}$
gives for the nonlinear E-J relation:
$~~E/(J R) = {\tilde E} {\bigl(}J/T^{1+2\nu}, L^{1/\nu}T{\bigr)} ~~$,
where $ E$ is a universal scaling function.
Adopting the accepted definition of the crossover current density,
$E/(J_x R) = 2$ yields: $~~J_x =T^{1+2\nu} f{\bigl(}L^{1/\nu}T{\bigr)}~~$,
where $f$ is an other universal function. Ref. \cite{young97} finds $J_x$ to be 
a universal function of $L^{1/\nu}T$, with $\nu = 1$. To test this 
proposition, we also plotted $J_x/T^{1+2\nu}$ as a function of 
$L^{1/\nu}T$. However we found no universal dependence whatsoever.
This clearly eliminates the possibility of a $T=0$ fixed point governing
the freezing behaviour of the London model.
Thus, remarkably, regarding the freezing transition the gauge glass
and the realistic vortex simulations give qualitatively different results.

We now understand that structural (window) glasses and systems with quenched 
disorder often behave quite similarly \cite{wolynes}. Their glassy phase exhibits 
different aging phenomena \cite{youngbook}. Measuring the two time correlation 
functions \cite{youngbook} of vortices, and comparing to the predicted power law 
relaxation forms would be a constructive test of the Vortex Molasses.

A word on the appropriateness of models with strong screening. 
Ref.\cite{young97} recalls that in the analogous 3d XY model {\it both} the 
screening length $\lambda$ and correlation length $\xi$ diverge, when $T_c$ is 
approached from {\it below}. In the critical region the exponent of 
$\lambda$ is half of $\xi$'s, and hence close enough to $T_c$ the proper 
characterization of the system {\it should} involve strong screening. 
{\it Above} $T_c$ the screening length on macroscopic scales is infinite, 
as we are in the Vortex Liquid. On the scale of intervortex separation, 
however, $\lambda$ equals its bare value. In general, the presence of the
other vortices generates a renormalized, scale dependent $\lambda(x)$.
Whether the model is in the strong or weak screening limit, will then
be determined by $\lambda(\xi)$ being greater or smaller than $\xi$.
By invoking that the critical behaviour around $T_c$ is typically symmetric, 
and that below $T_c$ we {\it are} in the strong screening limit, we expect
$\lambda(\xi) < \xi$, i.e. the screening remaining essential for understanding
the physics of the model, the starting point of our simulations.

In conclusion, we collected several numerical and experimental
results, which are at odds with the Vortex Glass theory of the 
freezing of the disordered vortex matter. 
Previous confirmations of the VG theory were reinterpreted
in terms of twin boundaries and proper accounts of the screening.
To reinvestigate the issue, 
we performed careful numerical simulations of the overdamped London - Langevin 
model, and used finite size scaling to analyze the data. We found that 
upon approaching
the transition the initial Vortex Glass type criticality is arrested
at some crossover temperature, where the vortex correlation length
catches up with the screening length. Below this temperature the timescale
continues growing very quickly, consistent with the Vogel-Fulcher form, 
while the spatial correlation length $\xi$ stops exhibiting any 
observable divergence. We call this mode of freezing the Vortex Molasses
scenario.

We thank G. Blatter, D. Fisher, A. Kapitulnik, P. Kes, W. Kwok, D. Lopez, 
V. Vinokur, A. Sudb{\o}, R.T. Scalettar, D. Nelson 
and especially D. Huse and P. Young for useful discussions. 
GTZ gratefully acknowledges the hospitality of the Argonne National Laboratories. 
This research was supported by NSF grant DMR 95-28535, and by a CLC and CULAR 
grants administered by the University of California.

\end{narrowtext}
\end{document}